\documentclass[reprint, amsmath, amssymb, aps, nofootinbib]{revtex4-2}

\usepackage{graphicx}
\usepackage{dcolumn}
\usepackage{bm}
\usepackage{braket}
\usepackage{amsmath}
\usepackage{here}
\usepackage{color}

\usepackage[small]{caption}
\usepackage[subrefformat=parens]{subcaption}
\usepackage{graphicx} 

\newcommand{\Slash}[1]{{\ooalign{\hfil/\hfil\crcr$#1$}}}
\def\Slash#1{\not\!\!#1}

\usepackage{color}

\begin{document}

\title{SO(3) real algebra method for SU(3) QCD at finite baryon-number densities} 

\author{Hideo~Suganuma}
\author{Kei Tohme}
\affiliation{
Department of Physics, Kyoto University, 
Kitashirakawaoiwake, Sakyo, Kyoto 606-8502, Japan\\
\email{suganuma@scphys.kyoto-u.ac.jp}
}

\date{\today}
\begin{abstract}
For SU(3) lattice QCD calculations at finite baryon-number densities, 
we propose the ``SO(3) real algebra method'', in 
which the SU(3) gauge variable is divided into the SO(3) and SU(3)/SO(3) parts.
In this method, we introduce the ``maximal SO(3) gauge'' 
by minimizing the SU(3)/SO(3) part of the SU(3) gauge variable.
In the Monte Carlo calculation, 
the SO(3) real algebra method employs  
the SO(3) fermionic determinant, 
i.e., the fermionic determinant 
of the SO(3) part of the SU(3) gauge variable, in the maximal SO(3) gauge, 
as well as the positive SU(3) gauge action factor $e^{-S_G}$.
Here, the SO(3) fermionic determinant is real, and it is non-negative 
for the even-number flavor case ($N_f=2n$) of the same quark mass, 
e.g., $m_u=m_d$.
%
%
The SO(3) real algebra method alternates between 
the maximal SO(3) gauge fixing and 
Monte Carlo updates on the SO(3) determinant and $e^{-S_G}$.
After the most importance sampling,
the ratio of the SU(3) and SO(3) fermionic determinants 
is treated as a weight factor.
If the phase factor of the ratio does 
not fluctuate significantly among the sampled gauge configurations 
for a set of parameters (e.g., volume, chemical potential, and quark mass), 
then SU(3) 
lattice QCD calculations at finite densities would be feasible.
\end{abstract}

\pacs{11.15.Ha, 12.12.38.Gc, 12.38.Mh}
\maketitle

\section{Introduction}

Quantum chromodynamics (QCD) has been established as the fundamental theory of the strong interaction 
since the field-theoretical proof of asymptotic freedom in QCD in 1973 \cite{GW73,P73,W24},
In fact, perturbative QCD \cite{M09} successfully describes high-energy hadron reactions within the framework of the parton model \cite{BP69}.
However, solving QCD in the low-energy region is quite difficult due to its strong-coupling nature,
which leads to nonperturbative properties such as color confinement \cite{S23} and 
spontaneous chiral-symmetry breaking \cite{NJL61}. 

The lattice QCD formalism \cite{W74,KS75} was developed as a robust nonperturbative method based on Euclidean QCD, and M.~Creutz first performed lattice QCD Monte Carlo calculations around 1980 \cite{C79,C80}. 
The fermion degrees of freedom can be included 
using the hybrid Monte Carlo algorithm \cite{DKPR87}.
Finite temperature QCD can be also studied 
in lattice QCD by imposing temporal
(anti)periodicity \cite{R12}. 
Nowadays, lattice QCD is a powerful and reliable method for analyzing nonperturbative QCD. 
Numerous lattice QCD studies have clarified various aspects of nonperturbative QCD \cite{R12} and 
finite temperature QCD including the QCD transition   \cite{HOTQCD14}. 

Nevertheless, some aspects of nonperturbative QCD remain poorly understood due to the limitations of lattice QCD. 
One such example is SU(3) color QCD at finite baryon densities, because lattice QCD Monte Carlo calculations break down due to the sign problem \cite{N22}.

To begin with, we briefly review SU(3) Euclidean QCD 
and the current status of high density QCD.
QCD is an SU(3) gauge theory described by the quark field $\psi(x)$ and the gluon field, 
\begin{eqnarray}
A_\mu(x) =\sum_{a=1}^8 A_\mu^a(x)T^a \in {\rm su}(3), \quad A_\mu^a(x)\in {\bf R},
\end{eqnarray}
where $T^a$ ($a=1,2,..., 8$) denote the SU(3) generators on  the color degrees of freedom. 
In the Euclidean metric, 
the QCD action $S[A, \psi, \bar \psi]$ is 
written as
\begin{eqnarray}
S[A, \psi, \bar \psi]=S_G[A]
+\int d^4x ~\bar \psi (\Slash D+m) \psi 
\in {\bf R},
\end{eqnarray}
with 
the SU(3) covariant derivative 
$D_\mu \equiv \partial_\mu +igA_\mu$.
(In this paper, we use hermite $\gamma$-matrices of 
$\gamma_\mu^\dagger=\gamma_\mu$ in the Euclidean metric.)
The gauge action $S_G[A]$ 
only contains the gluon field strength 
$G_{\mu\nu}=G_{\mu\nu}^aT^a \in {\rm su}(3)$ with $G_{\mu\nu}^a \in {\bf R}$, 
and it is real and non-negative 
in the Euclidain metric:
\begin{eqnarray}
S_G[A] \equiv 
\int d^4x ~\frac12{\rm Tr}(G_{\mu\nu}G_{\mu\nu})
=  \int d^4x ~\frac14G_{\mu\nu}^aG_{\mu\nu}^a 
\cr \ge 0.
\end{eqnarray}

At zero density, the Euclidean QCD generating functional is expressed by 
\begin{eqnarray}
Z&=&\int DA D\psi D\bar \psi e^{-S[A, \psi, \bar \psi]} \cr
&=&\int DAe^{-S_G[A]}{\rm Det}(\Slash D+m),
\end{eqnarray}
where the gauge action factor is positive,  
$e^{-S[A]} > 0$,
and the (zero-density) fermionic determinant 
is real, 
\begin{eqnarray}
{\rm Det}(\Slash D+m) \in {\bf R}.
\end{eqnarray}
In fact, one can derive  
\begin{eqnarray}
&&[{\rm Det}(\Slash D+m)]^* 
={\rm Det}[(\Slash D+m)^\dagger] 
={\rm Det}(-\Slash D+m)\cr
&=&{\rm Det}[\gamma_5(-\Slash D+m)\gamma_5]
={\rm Det}(\Slash D+m),
\end{eqnarray}
using the anti-hermitian nature of the covariant derivative, 
$\Slash D^\dagger=-\Slash D$, 
and the anticommutable nature of 
$\gamma_\mu$ and $\gamma_5$.
Then, for the even-number flavor case of the same quark mass, 
the fermionic determinant is non-negative.
In this case, the gauge action factor and the fermionic determinant can be identified as probability densities 
in the Monte Carlo method,
and lattice QCD calculations can be performed at zero density. 
Lots of studies of the QCD vacuum and hadrons 
have been conducted using lattice QCD Monte Carlo calculations for both zero-temperature \cite{R12}
and finite-temperature systems \cite{HOTQCD14}.

It is also important to understand 
finite baryon-number density QCD 
and to clarify the QCD phase diagram \cite{YHM05},  
based on the fundamental principles of the strong interaction.  
The QCD phase diagram can include 
not only deconfinement and chiral restoration, 
but also various phases such as 
pion condensation\cite{M73,TTTT78}, 
color superconductivity \cite{ASRS08},
and quarkyonic matter \cite{MP07,MRS09}.
It is also applied to the interior region of nuclear systems, and neutron stars. 
However, a serious problem called the sign problem 
prevents practical calculations of 
SU(3) lattice QCD at finite densities \cite{N22}.
In lattice QCD, many methods have been proposed to overcome the sign problem and to clarify the QCD phase diagram with a quark chemical potential $\mu$ and temperature $T$.

For instance, the Taylor expansion method \cite{FP02,AEHKK02} employs a Taylor expansion 
with respect to $\mu/T$ and is effective  
in regions of small chemical potentials, 
such as $\mu < T$. 
The complex Langevin method \cite{PW81,P83,ASS10,S14} 
is one of the hopeful methods because its statistical sampling methods can handle complex interactions without encountering algorithmic difficulties and enable nonperturbative calculations, although 
the theoretical foundation for the convergence is incomplete and the correctness of solutions must be verified for each dynamical system \cite{NNS16,TD16}.
The reweighing method \cite{FS88,FS89} 
is a useful general approach to deal with 
controllable and difficult factors separately \cite{SOK24}
and has been applied to finite density QCD \cite{FK02}.
An analytic continuation has been attempted from  systems with pure imaginary chemical potentials $\mu\in i{\bf R}$ \cite{AKW99,DL03,NN11,BBFGKRS15}, 
where no sign problem appears.
The Lefschetz thimble method \cite{W11} 
and its applications have been investigated
as an interesting approach to high density QCD. 
This method uses a complex path where the imaginary part 
of the action is constant or optimized to control 
the sign problem \cite{ABB16,MKO17,MKO18,FU17,FM21}.
For heavy quark systems, a sign-problem-free effective model constructed by combining strong coupling and hopping parameter expansion has been investigated as an indirect approach to finite density QCD \cite{FLLP12}.
The complex phase of the fermionic determinant in finite density QCD has been analyzed in the large $N_c$ context \cite{HMY12}.
Despite the development of 
calculation techniques mentioned above, 
no conclusive method has yet been established. 

Instead, many authors have studied systems that resemble finite density QCD without the sign problem. 
For instance, since SU(2) is pseudo-real, 
SU(2) color QCD does not encounter the sign problem, 
and high-density systems have been studied in lattice QCD  \cite{HKLM99,BGIKM15,BBIKM18,IIL20,I25}, although baryons are bosons in SU(2) QCD.
Even in SU(3) QCD, 
no sign problem appears for the systems 
with isospin chemical potentials \cite{AKW99,SS01,KS02}
for the $N_f=2$ case with the same u, d quark mass,
and lattice QCD simulations have been performed  \cite{HOTQCD19}.
This type of system could be found in the mismatch 
of the proton-to-neutron ratio, as is the case inside neutron stars.
Also, in SU(3) QCD, no sign problem appears for systems with a chiral chemical potential \cite{Y11}
as a result of the chiral magnetic effect \cite{CME08}. 
Such a system may be realized in the presence 
of strong magnetic fields \cite{ST91,MS15,DMS10}.
Although these studies may provide insight into the qualitative behavior of high density QCD, these examples are rather special or differ significantly from real finite-density SU(3) color QCD.

In this paper, we propose a new method of the ``SO(3) real algebra method'' for lattice QCD calculations at finite 
baryon-number density.
This paper is organized as follows. 
Section II provides a brief review of finite density QCD and the sign problem, a significant difficulty associated with Monte Carlo calculations.
In Sec.~III, we investigate the SO(3) subalgebra/subgroup of the SU(3) algebra/group, as well as the SU(3)/SO(3) coset space. 
In Sec.~IV, we present factorization of an SU(3) group element into an SO(3) element and an SU(3)/SO(3) element, which results in an SO(3) projection. 
In Sec.~V, we introduce a new gauge of the ``maximal SO(3) gauge'' in SU(3) QCD.
In Sec.~VI, we formulate the maximal SO(3) gauge in SU(3) lattice QCD.
In Sec.~VII, we present the formalism of the ``SO(3) real algebra method'' for continuum QCD at finite baryon-number density.
In Sec.~VIII, we formulate the SO(3) real algebra method in SU(3) lattice QCD.
Section IX is devoted to a summary and conclusion.

\section{Finite baryon-number density QCD}

In this section, we briefly review finite density SU(3) QCD 
with a quark-number chemical potential $\mu$, 
which is equal to one third of the baryon-number chemical potential.
In the physical case, the
chemical potential $\mu$ is real, $\mu \in {\bf R}$.
However, here it is treated as a general complex number, $\mu\in{\bf C}$.

The origin of the sign problem lies in the complex value of the action or the fermionic determinant for SU(3) QCD at finite densities, even in the Euclidean metric.
In the Euclidean metric, the finite-density SU(3) QCD action $S[A, \psi, \bar \psi;\mu]$ with the quark-number 
chemical potential $\mu$ is generally complex, 
\begin{eqnarray}
S[A, \psi, \bar \psi;\mu]=S_G[A]
+\int d^4x ~\bar \psi (\Slash D+m+\mu\gamma_4) \psi  \cr
\in {\bf C}.
\end{eqnarray}
Accordingly, the SU(3) fermionic determinant 
$\Delta_{\rm FD}[A](\mu)$ is generally complex 
in finite density QCD, 
\begin{eqnarray}
\Delta_{\rm FD}[A](\mu)\equiv
{\rm Det}(\Slash D+m+\mu\gamma_4) \in {\bf C},
\end{eqnarray}
after integrating out the fermion field in the QCD generating functional, 
\begin{eqnarray}
Z(\mu)&=&\int DA D\psi D\bar \psi e^{-S[A, \psi, \bar \psi;\mu]} \cr
&=&\int DAe^{-S_G[A]}{\rm Det}(\Slash D+m+\mu\gamma_4).
\end{eqnarray}
Therefore, in the case of finite densities,
unlike ordinary lattice QCD calculations, it is impossible to identify 
the action factor $e^{-S[A, \psi, \bar \psi;\mu]}\in {\bf C}$ or
the fermionic determinant $\Delta_{\rm FD}[A](\mu)\in {\bf C}$ as a probability density in the Monte Carlo calculation.  

Just as the partition function is real in a usual statistical system, the generating functional $Z(\mu)$ is also real.
In finite density SU(3) QCD, 
although $Z(\mu)$ itself is real, the fermionic determinant is generally complex, 
and its phase factor fluctuates violently and cancels out, making numerical simulations impossible. This is the sign problem.
However, if the phase factor is sufficiently suppressed during the most importance sampling, Monte Carlo simulations could be performed.
Note again that the generating functional $Z(\mu)$ itself is real, and therefore 
the sampled gauge configurations 
can, in principle, be chosen so that 
their determinant is close to being real.


To examine the complex conjugate of the fermionic determinant $\Delta_{\rm FD}[A](\mu)$ in finite density QCD, 
we use the Dirac representation for the Euclidean (hermite)  $\gamma$-matrices, 
in which only $\gamma_2$ is purely complex, $\gamma_2^*=-\gamma_2$. 
In the Dirac representation, 
the complex conjugation matrix $C$ is expressed as
\begin{eqnarray}
C \equiv i \gamma_5\gamma_2,
\end{eqnarray} 
and it satisfies 
\begin{eqnarray}
\gamma_\mu^*=C \gamma_\mu C^{-1} \quad (\mu=1,2,3,4).
\label{eq:CC}
\end{eqnarray}
The complex conjugate of the SU(3) fermionic determinant $\Delta_{\rm FD}[A](\mu)$ 
can be written as
\begin{eqnarray}
\Delta_{\rm FD}[A](\mu)^* 
&=&{\rm Det}\{\gamma_\alpha (\partial_\alpha +igA_\alpha)+m+\mu\gamma_4\}^* \cr
&=&{\rm Det}\{\gamma_\alpha^*(\partial_\alpha -ig A_\alpha^*)+m+\mu^*\gamma_4^*\} \cr
&=&{\rm Det}[C\{\gamma_\alpha (\partial_\alpha -ig A_\alpha^*)+m+\mu^*\gamma_4\}C^{-1}] \cr
&=&{\rm Det}\{\gamma_\alpha (\partial_\alpha -ig A_\alpha^*)+m+\mu^*\gamma_4\} \cr
&=&\Delta_{\rm FD}[-A^*](\mu^*).
\end{eqnarray}
In fact, there is an important relation of 
\begin{eqnarray}
\Delta_{\rm FD}[A](\mu)^*=\Delta_{\rm FD}[-A^*](\mu^*),
\end{eqnarray}
and it is generally complex in SU(3) QCD for the physical case of $\mu \in {\bf R}$. 

In this paper, we focus on the property of 
\begin{eqnarray}
-A_{\rm SO(3)}^\mu(x)^*=A_{\rm SO(3)}^\mu(x) 
\end{eqnarray}
for the SO(3) gauge field 
\begin{eqnarray}
A_{\rm SO(3)}^\mu(x)\equiv \sum_{i=2,5,7}A^\mu_i(x)T^i \in {\rm so}(3) \subset {\rm su}(3),
\label{eq:SO(3)gluon}
\end{eqnarray}
which is extracted from 
the SU(3) gauge field 
$A_\mu(x)=\sum_{a=1}^8 A_\mu^a(x)T^a \in {\rm su}(3)$
in QCD.
In fact, as far as the SO(3) gauge part
$A_{\rm SO(3)}$ is concerned, the fermionic determinant is real as
\begin{eqnarray}
\Delta_{\rm FD}[A_{\rm SO(3)}](\mu)^*&=&\Delta_{\rm FD}[-A_{\rm SO(3)}^*] (\mu^*) \cr
&=&\Delta_{\rm FD}[A_{\rm SO(3)}](\mu) 
\ \in {\bf R}
\end{eqnarray}
for the physical case of real finite densities, $\mu \in{\bf R}$. 

In this paper, 
we explore the possibility of suppressing the net phase factor of the SU(3) fermionic determinant of the sampled configurations by using 
a new gauge, the ``maximal SO(3) gauge'', which maximally forces the sampled SU(3)  gauge configurations 
to closely resemble real ones.

Toward lattice QCD calculations at finite densities, 
we propose a new method, the ``SO(3) real algebra method'', using the SO(3) part of SU(3) QCD in the maximal SO(3) gauge.

\section{SO(3) subalgebra/subgroup in SU(3) and SU(3)/SO(3) coset space}

For our new concepts of the ``maximal SO(3) gauge'' and the ``SO(3) real algebra method'', properties of SO(3) and SU(3)/SO(3) play an essential role.
In this section, after reviewing the SO(3) subalgebra of the SU(3) algebra,  
we investigate the SO(3) subgroup of the SU(3) group, 
as well as the SU(3)/SO(3) coset space.

\subsection{SO(3) subalgebra of SU(3) algebra}

We use the SU(3) algebra in the fundamental representation, where the SU(3) generators $T^a$ are expressed using 
the Gell-Mann matrices $\lambda^a$, 
\begin{eqnarray}
T^a=\frac{\lambda^a}{2} \quad (a=1, 2, 3, ... , 8).
\end{eqnarray}
The SU(3) generators $T^a$
satisfy the algebra relation 
\begin{eqnarray}
[T^a, T^b]=if^{abc}T^c
\label{eq:su3algebra}
\end{eqnarray}
with the structure constant $f^{abc} \in {\bf R}$.

The SU(3) algebra contains an SO(3) subalgebra. 
We divide the SU(3) generators $T^a (a=1,2,3,...,8)$ 
into the SO(3) generators 
$T^i (i=2,5,7)$ and the SU(3)/SO(3) generators $T^{\bar i} ({\bar i}=1,3,4,6,8)$.

In this paper, we denote 
the SO(3) generators by $T^i~(i \equiv 2,5,7)$, which are explicitly defined by 
\begin{eqnarray}
T^2&=&\frac12\begin{pmatrix}  
0 & -i & 0 \\
i & 0 & 0 \\
0 & 0 & 0
\end{pmatrix}, \quad
T^5=
\frac12\begin{pmatrix}
0 & 0 & -i \\
0 & 0 & 0 \\
i & 0 & 0
\end{pmatrix}, \cr
T^7&=&
\frac12\begin{pmatrix}  
0 & 0 & 0 \\
0 & 0 & -i \\
0 & i & 0
\end{pmatrix}.
\end{eqnarray}

On the other hand, we define the SU(3)/SO(3) generators as 
$T^{\bar i}~({\bar i} \equiv 1,3,4,6,8)$, 
where the indices are  barred, 
and they are explicitly defined by
\begin{eqnarray}
T^1&=&\frac12\begin{pmatrix}  
0 & 1 & 0 \\
1 & 0 & 0 \\
0 & 0 & 0
\end{pmatrix}, \quad
T^3=\frac12
\begin{pmatrix}  
1 & 0 & 0 \\
0 & -1 & 0 \\
0 & 0 & 0
\end{pmatrix}, \cr
T^4&=&\frac12
\begin{pmatrix}  
0 & 0 & 1 \\
0 & 0 & 0 \\
1 & 0 & 0
\end{pmatrix}, \quad
T^6=\frac12\begin{pmatrix}  
0 & 0 & 0 \\
0 & 0 & 1 \\
0 & 1 & 0
\end{pmatrix}, \cr
T^8&=&\frac1{2\sqrt3} 
\begin{pmatrix}  
1 & 0 & 0 \\
0 & 1 & 0 \\
0 & 0 & -2
\end{pmatrix}.
\end{eqnarray}
For simplicity, we fix the above-mentioned  representation in this paper.

In this representation, 
the SO(3) generators $T^i$ 
satisfy 
\begin{eqnarray}
T^{i*}=~^tT^i=-T^i \quad (i=2,5,7),
\end{eqnarray}
and the SU(3)/SO(3) generators $T^{\bar{i}}$ 
satisfy
\begin{eqnarray}
T^{\bar{i}*}=~^tT^{\bar{i}}=T^{\bar{i}} \quad (\bar{i}=1,3,4,6,8).
\end{eqnarray}

Next, in terms of SO(3) and SU(3)/SO(3) generators, we consider 
the SU(3) structure constant 
$f^{abc} \in {\bf R}$ ($a,b,c=1,2...,8$) 
satisfying the SU(3) algebra relation (\ref{eq:su3algebra}).
The SU(3) structure constant $f^{abc}$ 
is a perfect anti-symmetric tensor, 
and its components are expressed as 
\begin{eqnarray}
f_{123}&=&1, \cr
f_{147}&=&-f_{156}=f_{246}=f_{257}=f_{345}=-f_{367}=\frac{1}{2}, \cr
f_{458}&=&f_{678}=\frac{\sqrt{3}}{2}.
\label{eq:strconst}
\end{eqnarray}
Regarding the structure constant related to the SO(3) algebra, we note the useful relation 
\begin{eqnarray}
f^{\bar{i}ij}
=-2i{\rm Tr}(T^{\bar{i}}[T^i,T^j])
&=&0 \cr
({\bar i}=1,3,4,6,8 &;& i,j=2,5,7)
\label{eq:strconst_so3}
\end{eqnarray}
because $[T^i,T^j]$ is an SO(3) algebra.
From the structure constant (\ref{eq:strconst}),
we also find 
\begin{eqnarray}
f^{\bar{i}\bar{j}\bar{k}}=0 \quad 
({\bar i}, {\bar j}, {\bar k}=1,3,4,6,8)
\label{eq:strc_su3/so3}
\end{eqnarray}
for the SU(3)/SO(3) generators.

In this paper, we define the ``SO(3) index'' 
as $i \equiv 2, 5, 7$, and we use 
the symbol $i$ to represent it.
Similarly, we define the ``SU(3)/SO(3) index'' 
as ${\bar i} \equiv 1, 3, 4, 6, 8$, 
and we use the anti-index symbol ${\bar i}$
to represent it.
Additionally, repeated indices of $i$ or ${\bar i}$ of mean the summation over SO(3) or SU(3)/SO(3) indices, respectively.

\subsection{SO(3) subgroup and SU(3)/SO(3) coset space in SU(3) group}

In this subsection, we investigate 
the SO(3) subgroup and the SU(3)/SO(3) coset space
in the SU(3) group.

Using three real parameters $\theta^i$, 
the SO(3) group element $u$ is expressed as
\begin{eqnarray}
u=e^{i\theta^iT^i}\in{\rm SO(3)},
\quad
\theta^i\in {\bf R}~~(i=2,5,7).
\end{eqnarray}
As an important property, 
the SO(3) group element $u$ is real, 
\begin{eqnarray}
u^*=u, \quad {\rm i.e.,}~~
u^\dagger=~^t u,
\end{eqnarray}
because of 
$
(i\theta^iT^i)^*=i\theta^iT^i.
$

The representative element of 
the SU(3)/SO(3) group element $M$ is 
chosen to be 
\begin{eqnarray}
M=e^{i\theta^{\bar{i}}T^{\bar{i}}}
&\in& {\rm SU(3)/SO(3)}, \cr 
&&\theta^{\bar {i}}\in {\bf R}~~ (\bar{i}=1,3,4,6,8)
\end{eqnarray}
using five real parameters $\theta^{\bar {i}}$.
The SU(3)/SO(3) group element $M$ is 
a symmetric matrix,
\begin{eqnarray}
~^tM=M, \quad {\rm i.e.,}~~ M^\dagger
=M^*,
\end{eqnarray}
because of 
$~^t(\theta^{\bar{i}}T^{\bar{i}})=\theta^{\bar{i}}T^{\bar{i}}$ and 
$~^t[(\theta^{\bar{i}}T^{\bar{i}})^n]=
(\theta^{\bar{i}}T^{\bar{i}})^n$ for $n=0,1,2,...$

For any symmetric matrix ${\cal S}$ and 
any anti-symmetric matrix ${\cal A}$, 
the trace of their product is zero, 
${\rm Tr}({\cal SA})=0$:
\begin{eqnarray}
{\rm Tr}({\cal SA})={\cal S}_{ij}{\cal A}_{ji}={\cal S}_{ji}(-{\cal A}_{ij})=-{\rm Tr}({\cal SA})=0.
\end{eqnarray}
Therefore, the SU(3)/SO(3) group element $M$ 
satisfies 
\begin{eqnarray}
{\rm Tr}(M T^i)=0 \quad (i=2,5,7)
\end{eqnarray}
because $M$ 
is symmetric and the SO(3) generators $T^i$ ($i=2,5,7$) are anti-symmetric.

\subsection{SO(3) subgroup element}

For the SO(3) subalgebra, we 
also use the alternative form of the generators $\Lambda^a$ ($a=1, 2, 3$) defined by  
\begin{eqnarray}
\Lambda^1 &\equiv& \lambda^7=
\begin{pmatrix}  
0 & 0 & 0 \\
0 & 0 & -i \\
0 & i & 0
\end{pmatrix}, \quad
\Lambda^2 \equiv -\lambda^5= 
\begin{pmatrix}  
0 & 0 & i \\
0 & 0 & 0 \\
-i & 0 & 0
\end{pmatrix},
\cr
\Lambda^3 &\equiv& \lambda^2= 
\begin{pmatrix}  
0 & -i & 0 \\
i & 0 & 0 \\
0 & 0 & 0
\end{pmatrix}.
\end{eqnarray}
The SO(3) generators $\Lambda^a~(a=1,2,3)$  are  expressed as 
\begin{eqnarray}
(\Lambda^a)_{bc}=-i\epsilon_{abc}
\end{eqnarray}
and satisfy 
\begin{eqnarray}
[\Lambda^a, \Lambda^b]=i\epsilon_{abc}\Lambda^c,
\end{eqnarray}
that is, the SU(2) algebra relation in the adjoint representation.

The general SO(3) algebra 
is expressed as 
\begin{eqnarray}
\theta \cdot \Lambda \equiv \theta^a\Lambda^a \in {\rm so}(3)
\end{eqnarray}
using three real parameters 
$\theta^i \in {\bf R}$.

We note the relation, 
\begin{eqnarray}
(\hat \theta \cdot \Lambda)^3
=\hat \theta \cdot \Lambda,
\end{eqnarray}
with $\hat \theta^i \equiv \theta^i/\theta$ and $\theta \equiv \sqrt{\theta^i\theta^i}$.
In fact, the relations 
\begin{eqnarray}
(\Lambda^a \Lambda^b)_{ik}=&&\!\!\!\!\!\!\!\!
\Lambda^a_{ij} \Lambda^b_{jk}=
-\epsilon_{aij}\epsilon_{bjk}=
\delta_{ab}\delta_{ik}-\delta_{ak}\delta_{bi},\\
(\Lambda^a \Lambda^b\Lambda^c)_{il}&=&
(\Lambda^a \Lambda^b)_{ik}\Lambda^c_{kl}=
-i(\delta_{ab}\delta_{ik}-\delta_{ak}\delta_{bi})\epsilon_{ckl}\cr
&=&-i(\delta_{ab}\epsilon_{cil}-\delta_{bi}\epsilon_{cal})
\end{eqnarray}
lead to 
\begin{eqnarray}
[(\hat \theta \cdot \Lambda)^3]_{il}&=&
\hat \theta^a \hat \theta^b \hat \theta^c
(\Lambda^a\Lambda^b\Lambda^c)_{il} \cr
&=&-i\hat \theta^a \hat \theta^b \hat \theta^c
(\delta_{ab}\epsilon_{cil}-\delta_{bi}\epsilon_{cal})\cr
&=&-i\hat \theta^c \epsilon_{cil}=\hat \theta^c
\Lambda^c_{il}=(\hat \theta \cdot \Lambda)_{il}.
\end{eqnarray}
Then, the SO(3) group element $u$ can be expressed as
\begin{eqnarray}
u &\equiv& e^{i\theta^a\Lambda^a} 
=e^{i\theta(\hat\theta\cdot \Lambda)} 
=\sum_{n=0}^\infty \frac{(i\theta)^n}{n!}(\hat \theta\cdot \Lambda)^n\cr
&=&1+i(\hat \theta \cdot \Lambda) \sin \theta
+(\hat \theta \cdot \Lambda)^2(\cos\theta-1).
\end{eqnarray}

In particle physics, for example, 
the SO(3) subgroup of SU(3) is used to describe the H dibaryon, i.e., H(uuddss), in the Skyrme model \cite{BBLRS84,BLRS85,JK85,SS98} 
and in holographic QCD \cite{MNS17,SM17} 
for the three-flavor case.

\section{Factorization of SU(3) group into SO(3) and SU(3)/SO(3) ; SO(3) projection}

In this section, we consider 
factorization of an SU(3) group into 
an SO(3) subgroup and an SU(3)/SO(3) coset space. 
Using this factorization, 
we also present an SO(3) projection of an SU(3) group element.

\subsection{Factorization of SU(3) group into SO(3) and SU(3)/SO(3)}

In this subsection, we introduce 
factorization of an SU(3) group element $U$ into 
an SO(3) element $u$ and an SU(3)/SO(3) element $M$. 
Group factorization generally has some uncertainty, reflecting the non-commutative property.
For instance, there are two typical types of factorization: $U=Mu$ and $U=uM$. 
However, once the method of factorization is determined, the procedure becomes practically unique. 
(Similar uncertainty appears in 
the factorization of the ${\rm SU}(N_c)$ 
link variable into ${\rm U}(1)^{N_c-1}$ and
${\rm SU}(N_c)/{\rm U}(1)^{N_c-1}$ 
variables
in the maximally Abelian gauge
in SU($N_c$) QCD \cite{IS00}. Also, 
similar uncertainty appears in 
the product Ansatz used to describe
the baryon-baryon interaction 
in the Skyrme model \cite{JJP85,YA85}.)

In this paper, we basically use factorization of an SU(3) group element $U$ into an SO(3) subgroup element 
$u$ and an SU(3)/SO(3) element $M$ as follow: 
\begin{eqnarray}
U=Mu \in {\rm SU}(3)
\end{eqnarray}
of which $u$ and $M$ take the definite form of 
\begin{eqnarray}
u&=& e^{i\theta^iT^i} \in {\rm SO(3)}~~~~~~~~~~~ (i=2,5,7), \cr M&=&e^{i\theta^{\bar{i}}T^{\bar{i}}} \in {\rm SU(3)/SO(3)}~~  (\bar{i}=1,3,4,6,8). 
\end{eqnarray}

To extract $M$ from $U$, we note 
the useful relation  
\begin{eqnarray}
U^{t}U=Mu \ ^{t}u^{t}M=Mu u^{\dagger}M=M^2 \in {\rm SU(3)}.
\label{eq:UtU}
\end{eqnarray}
Based on this relation, we consider the uniqueness of the factorization.
In the case of $U=Mu=M'u'$, 
one finds 
$U^tU=M^2=M'^2 \in {\rm SU(3)}$.    
If $M \sim 1$, by expanding 
$M=e^{i\theta^{\bar{i}}T^{\bar{i}}}=1+i\theta^{\bar{i}}T^{\bar{i}}+\cdots$, 
$M^2=M'^2$ yields $M=M'$, 
which leads to 
$u=M^{\dagger}U=M'^{\dagger}U=u'$, 
and therefore the factorization 
is found to be unique.
Note also that, unlike SU(2), $-1$ is not an element of the SU(3) group.

\subsection{SO(3) projection for SU(3) group element}

In this subsection, 
we introduce the ``SO(3) projection'' 
for an SU(3) group element $U$.
For the factorization $U=Mu$,
we define the SO(3) projection as the replacement of 
\begin{eqnarray}
U=Mu \in {\rm SU(3)} ~~\rightarrow~~ u \in {\rm SO(3)}.
\end{eqnarray}

\section{Maximal SO(3) gauge in SU(3) QCD}

In this section, 
we introduce a new gauge, which we call the ``maximal SO(3) gauge'', in QCD of an SU(3) gauge theory, 
mainly in the Euclidean metric.
We divide the gluon field $A_\mu(x)=A_\mu^a(x)T^a \in {\rm su(3)}$ into the SO(3) part and the SU(3)/SO(3) part, 
\begin{eqnarray}
A_\mu(x)=A_\mu^{\rm SO(3)}(x)+A_\mu^{\rm SU(3)/SO(3)}(x) 
\in {\rm su(3)}, ~~~~~
\cr\cr
\left\{
 \begin{aligned}
&A_\mu^{\rm SO(3)}(x)\equiv \sum_{i=2,5,7} A_\mu^i(x)T^i \in {\rm so}(3) \in {\rm su(3)}\cr
&A_\mu^{\rm SU(3)/SO(3)}(x)\equiv
\sum_{{\bar{i}}=1,2,4,6,8} A_\mu^{\bar{i}}(x)T^{\bar{i}} \in {\rm su(3)}. 
 \end{aligned}
\right.
\end{eqnarray}

\subsection{Definition}

In Euclidean SU(3) QCD, 
the maximal SO(3) gauge is defined to minimize 
\begin{eqnarray}
R[A] &\equiv& \int d^4x \sum_{\mu}\sum_{\bar{i}=1,3,4,6,8} A_\mu^{\bar{i}}(x)^2 \cr
&=&\int d^4x \sum_{\mu} \{A_\mu^1(x)^2+A_\mu^3(x)^2\cr 
&&~~~~~~~~~+A_\mu^4(x)^2 
+A_\mu^6(x)^2 +A_\mu^8(x)^2\} 
\label{eq:MSO3}
\end{eqnarray}
by SU(3) gauge transformation.
The functional $R[A]$ is also expressed as 
\begin{eqnarray}
R[A] 
&=& 4 \int d^4x\sum_{\mu}\sum_{\bar{i}=1,3,4,6,8}[{\rm Tr}~(T^{\bar{i}}A_\mu(x))]^2 \cr
&=& 2 \int d^4x\sum_{\mu}~
[{\rm Tr}~A^{\rm SU(3)/SO(3)}_\mu(x)^2],
\end{eqnarray}
using the ``SU(3)/SO(3) part''
defined by 
\begin{eqnarray}
A_{\rm SU(3)/SO(3)}^\mu(x)\equiv\sum_{\bar{i}=1,3,4,6,8} A^\mu_{\bar{i}}(x)T^{\bar{i}} \in {\rm su(3)}. 
\label{eq:SU(3)/SO(3)gluon}
\end{eqnarray}
In fact, roughly speaking,  
the maximal SO(3) gauge globally minimizes
the SU(3)/SO(3) part, 
$A_{\rm SU(3)/SO(3)}^\mu(x)$, 
in the SU(3) gauge field $A_\mu(x)=A_\mu^a(x) T^a\in{\rm su}(3)$.

\subsection{Residual gauge symmetry}

In the maximal SO(3) gauge,  
an SO(3) gauge symmetry remains, 
since $R[A]$ is invariant under 
an arbitrary SO(3) gauge transformation 
\begin{eqnarray}
A^\mu \rightarrow \Omega_{\rm SO(3)}\left(A^\mu+\frac1{ig}\partial^\mu\right) \Omega_{\rm SO(3)}^\dagger
\end{eqnarray}
with an SO(3) gauge function 
\begin{eqnarray}
\Omega_{\rm SO(3)}(x)=e^{i\chi^i(x)T^i}\in {\rm SO}(3), ~\chi^i \in {\bf R}~(i=2,5,7). 
\end{eqnarray}

The proof is as follows:
Here, we divide the SU(3) indices 
$a\equiv 1,2,..,8$ into 
$i \equiv 2,5,7$ and $\bar{i} \equiv 1,3,4,6,8$.
Since any SO(3) transformation is achieved by performing 
an infinite number of infinitesimal transformations, 
we consider infinitesimal SO(3) gauge transformation,
\begin{eqnarray}
\Omega_{\rm SO(3)}(x)=1+i\chi^i(x)T^i, \quad |\chi^i(x)| \ll 1,
\end{eqnarray}
up to the first order of $\chi^i(x)$.
By this transformation, 
the SU(3)/SO(3) gluon component 
$A^{\bar{i}}_\mu(x)$ is transformed as 
\begin{eqnarray}
&&A^{\bar{i}}_\mu(x)=2~{\rm Tr}~[T^{\bar{i}} A_\mu(x)]\cr
&\rightarrow& 
2~{\rm Tr}~\Big[T^{\bar{i}}~
\Omega_{\rm SO(3)}
\Big(A_{\mu}+\frac{1}{ig}\partial_\mu\Big)
\Omega^{\dagger}_{\rm SO(3)}
\Big]
\cr
&=&
2~{\rm Tr}~[T^{\bar{i}}(1+iT^i\chi^i)A_{\mu}(1-iT^j\chi^j)]
\cr
&& \qquad -
\frac{2}{g}{\rm Tr}[T^{\bar{i}}\partial_\mu(T^i\chi^i)] \cr
&=&A_\mu^{\bar{i}}(x)
+2~{\rm Tr}(iT^i\chi^i[A_\mu, T^{\bar i}]) \cr
&=&A_\mu^{\bar{i}}(x)+\delta A_\mu^{\bar{i}}(x),
\end{eqnarray}
where the term including a derivative 
is zero because of  
${\rm Tr}(T^{\bar{i}}T^i)=0$.
Then, the variation of the SU(3)/SO(3) gluon component  $A^{\bar{i}}_\mu(x)$ 
is given as 
\begin{eqnarray}
\delta A_\mu^{\bar{i}}(x)
&\equiv& 2{\rm Tr}(iT^i\chi^i[A_\mu, T^{\bar i}])
=2i\chi^iA_\mu^a{\rm Tr}(T^i[T^a, T^{\bar i}]) \cr
&=&-2\chi^iA_\mu^a f^{a\bar{i}b}{\rm Tr}(T^iT^b)
=f^{i\bar{i}a} \chi^iA_\mu^a  \cr
&=&f^{i\bar{i}j} \chi^iA_\mu^j 
+f^{i\bar{i}\bar{j}} \chi^iA_\mu^{\bar{j}} 
=f^{i\bar{i}\bar{j}} \chi^iA_\mu^{\bar{j}},
\end{eqnarray}
because of 
$f^{\bar{i}ij}=0$, 
as was shown in Eq.(\ref{eq:strconst_so3}).
For the infinitesimal SO(3) gauge transformation,
the variation of $R[A]$ is written as
\begin{eqnarray}
\delta R[A] &\equiv& R[A+\delta A]-R[A]
=2\int d^4x~A_\mu^{\bar{i}}(x)~\delta  A_\mu^{\bar{i}}(x) \cr
&=&2\int d^4x~A_\mu^{\bar{i}}(x) 
\cdot f^{i\bar{i}\bar{j}} \chi^i(x)A_\mu^{\bar{j}}(x) \cr
&=&2\int d^4x~
\chi^i(x)
f^{i\bar{i}\bar{j}}
A_\mu^{\bar{i}}(x) 
A_\mu^{\bar{j}}(x)=0.
\end{eqnarray}
Since any SO(3) transformation is achieved through an
infinite number of infinitesimal transformations,
$R[A]$ is unchanged by any SO(3) gauge transformation.  

Thus, the maximal SO(3) gauge is a type of partial gauge fixing, and QCD in this gauge behaves as an SO(3) gauge theory.
(This is similar to the maximally Abelian gauge, which preserves 
Abelian gauge symmetry. 
QCD in this gauge behaves as an Abelian gauge theory \cite{IS00,KLSW87,KSW87,SY90,IS99,SS14,SS15}.)

\subsection{Local gauge-fixing condition}

Using the SO(3) covariant derivative
$D_{\rm SO(3)}^\mu$ defined by
\begin{eqnarray}
D_{\rm SO(3)}^\mu
\equiv \partial^\mu+igA_{\rm SO(3)}^\mu
= \partial^\mu+ig\sum_{i=2,5,7}A_i^\mu T^i,
\label{eq:SO3CD}
\end{eqnarray}
the local gauge-fixing condition of 
the maximal SO(3) gauge is expressed as
\begin{eqnarray}
\left[D^\mu_{\rm SO(3)},A
_\mu^{\rm SU(3)/SO(3)}\right]=0,
\label{eq:localGF}
\end{eqnarray}
or equivalently, 
\begin{eqnarray}
\Big[\partial^\mu+ig\sum_{i=2,5,7}A^\mu_i T^i, \sum_{{\bar i}=1,3,4,6,8} A_\mu^{\bar{i}}T^{\bar{i}}\Big]=0.
\label{eq:localGF1}
\end{eqnarray}
Here, $A_{\rm SO(3)}^\mu(x)$ and 
$A^\mu_{\rm SU(3)/SO(3)}(x)$ are 
the SO(3) part 
$A_{\rm SO(3)}^\mu(x)\equiv \sum_{i=2,5,7}A^\mu_i(x)T^i$ 
and the SU(3)/SO(3) part 
$A_{\rm SU(3)/SO(3)}^\mu(x)\equiv\sum_{\bar{i}=1,3,4,6,8} A^\mu_{\bar{i}}(x)T^{\bar{i}}$  
of the gluon field $A^\mu(x)\in {\rm su(3)}$, 
respectively.
According to the SO(3) covariant derivative,  
the SO(3) gauge symmetry is found to remain 
in this local gauge-fixing condition.

Using the SO(3) index $i$ 
and SU(3)/SO(3) index ${\bar i}$,
the local gauge-fixing condition 
(\ref{eq:localGF1}) is expressed as 
\begin{eqnarray}
\partial^\mu A_\mu^{\bar{i}}
+gf^{i\bar{i}\bar{j}}A^\mu_i A_\mu^{\bar{j}}=0.
\label{eq:localGF2}
\end{eqnarray}
Here, we have used the algebra relation, 
\begin{eqnarray}
    [T^i, T^{\bar i}]=if^{i\bar{i}a}T^a
=if^{i\bar{i}j}T^j+if^{i\bar{i}\bar{j}}T^{\bar{j}}
=if^{i\bar{i}\bar{j}}T^{\bar{j}},
\end{eqnarray}
due to $f^{\bar{i}ij}=0$ in Eq.~(\ref{eq:strconst_so3}). 

The derivation of the local gauge-fixing condition (\ref{eq:localGF}) is as follows.
As was shown in the previous subsection,
the maximal SO(3) gauge is SO(3) gauge invariant; therefore,  
it is sufficient to consider 
the SU(3)/SO(3) gauge transformation.
We consider infinitesimal SU(3)/SO(3) gauge transformation, 
\begin{eqnarray}
\Omega_{\rm SU(3)/SO(3)}(x)&=&e^{i\chi^{\bar i}(x)T^{\bar i}}
\cr
&\simeq& 1+i\chi^{\bar i}(x)T^{\bar i}, 
\quad |\chi^{\bar i}(x)| \ll 1,~~~
\end{eqnarray}
up to the first order of $\chi^{\bar i}(x)$.
By this transformation, 
the SU(3)/SO(3) gluon component 
$A^{\bar{i}}_\mu(x)$ is transformed as 
\begin{eqnarray}
&&A^{\bar{i}}_\mu(x)=2~{\rm Tr}~[T^{\bar{i}} A_\mu(x)]\cr
&\rightarrow& 
2~{\rm Tr}~\Big[T^{\bar{i}}~
\Omega_{\rm SU(3)/SO(3)}
\Big(A_{\mu}+\frac{1}{ig}\partial_\mu\Big)
\Omega^{\dagger}_{\rm SU(3)/SO(3)}
\Big]
\cr
&=&2~{\rm Tr}~[T^{\bar{i}}(1+iT^{\bar j}\chi^{\bar j})A_{\mu}(1-iT^{\bar k}\chi^{\bar k})] 
\cr
&& \qquad -
\frac{2}{g}{\rm Tr}[T^{\bar{i}}\partial_\mu(T^{\bar k}\chi^{\bar k})] \cr
&=&A_\mu^{\bar{i}}(x)
+2{\rm Tr}(iT^{\bar{j}}\chi^{\bar j}[A_\mu, T^{\bar i}]) 
-\frac{1}{g}\partial_\mu \chi^{\bar i} \cr
&=&A_\mu^{\bar{i}}(x)+\delta A_\mu^{\bar{i}}(x).
\end{eqnarray}
Then, the variation of the SU(3)/SO(3)
gluon component $A_\mu^{\bar i}(x)$ 
is expressed as
\begin{eqnarray}
\delta A_\mu^{\bar i}(x)&\equiv&2{\rm Tr}(iT^{\bar{j}}\chi^{\bar j}[A_\mu, T^{\bar i}]) 
-\frac{1}{g}\partial_\mu\chi^{\bar i} \cr
&=&2i\chi^{\bar j}A_\mu^a{\rm Tr}(T^{\bar{j}}[T^a, T^{\bar i}]) -\frac{1}{g}\partial_\mu\chi^{\bar i}
\cr
&=&
-\chi^{\bar j}A_\mu^a
f^{a\bar{i}\bar{j}}-\frac{1}{g}\partial_\mu\chi^{\bar i} \cr
&=&
-\chi^{\bar j}A_\mu^i
f^{i\bar{i}\bar{j}}
-\frac{1}{g}\partial_\mu\chi^{\bar i}
\end{eqnarray}
because of $f^{\bar{i}\bar{j}\bar{k}}=0$ in Eq.(\ref{eq:strc_su3/so3}).
Then, one finds 
\begin{eqnarray}
A_\mu^{\bar i}(x)\delta A_\mu^{\bar i}(x)
&=&-A_\mu^{\bar i}(
\frac{1}{g}\partial_\mu\chi^{\bar i}+\chi^{\bar j}A_\mu^i
f^{i\bar{i}\bar{j}})\cr
&=&
-\frac{1}{g}\partial_\mu\chi^{\bar i}
\cdot A_\mu^{\bar i}
+\chi^{\bar i}f^{i\bar{i}\bar{j}}
A_\mu^iA_\mu^{\bar j},~~~~ 
\end{eqnarray}
where the indices ${\bar i}$ and ${\bar j}$ are interchanged in part. 
For the infinitesimal SU(3) gauge transformation,
the variation of $R[A]$ is written as
\begin{eqnarray}
&&\delta R[A] \equiv R[A+\delta A]-R[A]
=2\int d^4x~A_\mu^{\bar{i}}(x)~\delta  A_\mu^{\bar{i}}(x) \cr
&&=2\int d^4x~
\Big(-\frac{1}{g}\partial_\mu\chi^{\bar i}
\cdot A_\mu^{\bar i}
+\chi^{\bar i}f^{i\bar{i}\bar{j}}
A_\mu^iA_\mu^{\bar j}\Big) \cr
&&=\frac{2}{g}\int d^4x~\chi^{\bar i}(x)
(\partial_\mu A_\mu^{\bar i}
+gf^{i\bar{i}\bar{j}}
A_\mu^iA_\mu^{\bar j}),
\end{eqnarray}
where a partial integration is performed with the boundary condition
that the gauge function vanishes at infinity, $\chi^{\bar i}(|x|=\infty)=0$.
The extreme condition of $R[A]$, $\delta R=0$, with respect to 
the infinitesimal gauge transformation with an arbitrary small $\chi^{\bar i}(x)$ leads to 
$\partial_\mu A_\mu^{\bar i}
+gf^{i\bar{i}\bar{j}}
A_\mu^iA_\mu^{\bar j}=0$, that is, 
the local gauge-fixing condition (\ref{eq:localGF2}).

\section{Maximal SO(3) gauge 
in SU(3) lattice QCD}

In this section, 
we formulate the ``maximal SO(3) gauge'' 
in SU(3) lattice QCD.

\subsection{Definition}

For the SU(3) link-variable $U_\mu(s)$, 
we consider the factorization of 
\begin{eqnarray}
U_\mu(s)=M_\mu(s)u_\mu(s) \in {\rm SU(3)}    
\end{eqnarray}
in terms of SO(3) and SU(3)/SO(3) variables,  
\begin{eqnarray}
u_\mu(s) &\equiv& e^{iT^i\theta_\mu^i(s)} \in {\rm SO(3)}, \cr
M_\mu(s) &\equiv& e^{iT^{\bar i}\theta_\mu^{\bar i}(s)} \in {\rm SU(3)/SO(3)}.
\end{eqnarray}
Due to 
$^tu_\mu(s)=u_\mu^\dagger(s)$ and $^tM_\mu(s)=M_\mu(s)$,
we find
\begin{eqnarray}
U_\mu(s)^tU_\mu(s)=M_\mu(s)^2,
\end{eqnarray}
which is the lattice QCD version of Eq.~(\ref{eq:UtU}).

In Euclidean SU(3) lattice QCD, 
the maximal SO(3) gauge 
is defined to maximize 
\begin{eqnarray}
R_L[U] &\equiv& \sum_{s,\mu} {\rm Re}~{\rm Tr}\{U_\mu(s)^tU_\mu(s)\} \cr
&=&
\sum_{s,\mu} {\rm Re}~{\rm Tr}\{M_\mu(s)^2\}
\label{eq:MSO3L}
\end{eqnarray}
by SU(3) gauge transformation.

Thus, in the maximal SO(3) gauge,  
the SU(3)/SO(3) factor $M_\mu(s)$ 
in the link-variable $U_\mu(s) \in {\rm SU(3)}$
is maximally close to unity, 
since $-1$ is not an SU(3) group element.
Therefore, one roughly expects 
\begin{eqnarray}
M_\mu(s) \sim 1,~~{\rm i.e.,}~~U_\mu(s)\sim u_\mu(s) 
\end{eqnarray}
in the maximal SO(3) gauge.

\subsection{Continuum limit}

We examine the continuum limit of 
the lattice QCD definition 
of the maximal SO(3) gauge.
In the maximal SO(3) gauge, 
$M_\mu(s)=e^{iT^{\bar{i}}\theta_\mu^{\bar i}(s)} \in {\rm SU(3)/SO(3)}$ is 
close to unity,  
and $\theta_\mu^{\bar i}(s)$ 
is considered to be small, 
which allows the following expansion:  
\begin{eqnarray}
U_\mu(s)^tU_\mu(s)&=&M_\mu(s)^2
=e^{2iT^{\bar{i}}\theta_\mu^{\bar{i}}(s)} \cr
&\simeq& 1+2iT^{\bar{i}}\theta_\mu^{\bar{i}}(s)-
2T^{\bar{i}}\theta_\mu^{\bar{i}}(s)T^{\bar{j}}\theta_\mu^{\bar{j}}(s),~~~~
\end{eqnarray}
and then
\begin{eqnarray}
R_L[U] &\equiv&
\sum_{s,\mu} {\rm Re~Tr}~\{U_\mu(s)^tU_\mu(s)\} \cr
&\simeq& -\sum_{s,\mu} {\rm Re~Tr}~
\theta_\mu^{\bar{i}}(s)\theta_\mu^{\bar{i}}(s)+{\rm const.}
\end{eqnarray}
Therefore, in the continuum limit, maximizing $R_L[U]$ 
corresponds to minimizing 
$R[A]$ defined in Eq.(\ref{eq:MSO3}).  
This yields the definition of 
the maximal SO(3) gauge in 
continuum SU(3) QCD, presented in Sec.~V.

\subsection{Residual gauge symmetry}

In the maximal SO(3) gauge,  
an SO(3) gauge symmetry remains, 
since $R_L[U]$ is invariant under 
an arbitrary SO(3) gauge transformation 
\begin{eqnarray}
U_\mu(s) \rightarrow v (s)U_\mu(s)v^\dagger(s+\hat \mu),
\end{eqnarray}
with an SO(3) gauge function $v(s) \in {\rm SO(3)}$.
In fact, by the SO(3) gauge transformation, 
$R_L[U]$ is unchanged as
\begin{eqnarray}
R_L[U] &\rightarrow& 
\sum_{s,\mu} {\rm Re}~{\rm Tr}\{
 v(s)U_\mu(s)v^\dagger(s+\hat \mu) \cr
 &&~~~~~~~~~~~\cdot
 ^tv^\dagger(s+\mu)^tU_\mu(s)^tv(s)\} \cr
&=&
\sum_{s,\mu} {\rm Re}~{\rm Tr}\{U_\mu(s)^tU_\mu(s)\}
=R_L[U],
\end{eqnarray}
because of the SO(3) property, $^tv(s)=v^\dagger(s)$.

For $U_\mu(s)=M_\mu(s)u_\mu(s)$, 
we consider the SO(3) gauge transformation 
of $M_\mu(s) \in {\rm SU(3)/SO(3)} $ and 
$u_\mu(s) \in {\rm SO(3)}$.
To keep the factorial form, we find 
\begin{eqnarray}
M_\mu(s)&\rightarrow& M_\mu^v(s) =v(s) M_\mu(s) v^\dagger(s) \in {\rm SU(3)/SO(3)}, \cr
u_\mu(s)&\rightarrow& u_\mu^v(s)=v(s) u_\mu(s) v^\dagger(s+\hat \mu) \in {\rm SO(3)}.
\end{eqnarray}
In fact, the SU(3)/SO(3) property of 
$^tM(s)=M(s)$ is kept by the SO(3) gauge transformation as  
\begin{eqnarray}
^t M^v(s)&=&^t\{v(s) M_\mu(s) v^\dagger(s)\} \cr
&=&
v(s) M_\mu(s) v^\dagger(s)=M^v_\mu(s)
\end{eqnarray}
because of $^tv(s)=v(s)^\dagger \in{\rm  SO(3)}$.

Thus, by the SO(3) gauge transformation,  
$M_\mu(s) \in {\rm SU(3)/SO(3)}$ obeys 
a unitary transformation at the same spacetime, and 
$u_\mu(s) \in {\rm SO(3)}$ behaves as a gauge variable that links neighboring spacetime sites.
(This situation is similar to 
the behavior of 
diagonal and off-diagonal gluons 
by residual Abelian gauge transformation 
in the maximally Abelian gauge 
\cite{IS00,IS99,SS14,SS15}.)

\subsection{SU(3) gauge transformation property}

Now, we consider general
SU(3) gauge transformation 
of the 
quantity $R_L[U]$ defined in Eq.(\ref{eq:MSO3L}).
We factorize the gauge function 
$V(s) \in {\rm SU(3)}$ 
as
\begin{eqnarray}
V(s) &=& v(s)\bar V(s) \quad {\rm or}
\quad
V^\dagger(s) = \bar V^\dagger (s) v^\dagger (s),
\cr
v(s) &\in& {\rm SO(3)}, \quad 
\bar V(s) \in {\rm SU(3)/SO(3)},
\end{eqnarray}
which satisfy 
$^tv_\mu(s)=v_\mu^\dagger(s)$ and $^t\bar V_\mu(s)=\bar V_\mu(s)$.

Then, the SU(3) link variable $U_\mu(s)$ 
is gauge transformed as  
\begin{eqnarray}
U_\mu(s) &\rightarrow& V(s)U_\mu(s)V^\dagger(s+\hat \mu) \cr
&=&v(s)\bar V(s) 
U_\mu(s)
\bar V^\dagger (s+\hat \mu) v^\dagger (s+\hat \mu), 
\end{eqnarray}
and then $R_L[U]$ is transformed as 
\begin{eqnarray}
&&R_L[U] \equiv \sum_{s,\mu} {\rm Re}~{\rm Tr}\{U_\mu(s)^tU_\mu(s)\} \cr
&\rightarrow&
\sum_{s,\mu} {\rm Re}~{\rm Tr}\{
\bar V(s)^2~
U_\mu(s)~
\bar V^{\dagger} (s+\hat \mu)^2 \
^tU_\mu(s)\}.~~~~~
\end{eqnarray}
Since the gauge transformation of $R_L[U]$ is 
independent of the SO(3) gauge function $v(s)$, we find again that 
the SO(3) part remains as a residual gauge symmetry.

\subsection{Local SU(3) gauge transformation}

For $R_L[U]$ defined in Eq.~(\ref{eq:MSO3L}), 
we examine the local SU(3) gauge transformation consisting of a nontrivial transformation only
at one site $s$ in lattice QCD,
\begin{eqnarray}
V(s')=V\delta_{s's}, \quad V\in {\rm SU(3)}.
\end{eqnarray}
As a result of this gauge transformation, 
$R_L[U]$ differs by $\Delta R_L$: 
\begin{eqnarray}
&&\Delta R_L \equiv  R_L[U^V]-R_L[U] \cr
&=&
\sum_{\mu=1}^4 
{\rm Re Tr}\{VU_\mu(s)^tU_\mu(s)^tV\} \cr
&+&
\sum_{\mu=1}^4 
{\rm Re Tr}\{U_\mu(s-\hat \mu)V^{\dagger}~^t V^{\dagger t} U_\mu(s-\hat \mu)\} \cr
&=&
{\rm Re Tr}\{^tVV
\sum_{\mu=1}^4 \sum_{\pm} U_{\pm \mu}(s)^tU_{\pm \mu}(s)\}. 
\end{eqnarray}
Here, we have used $U_{-\mu}(s)\equiv U_\mu(s-\mu)^\dagger$ and 
${\rm Re Tr}(M)={\rm Re Tr}(M^\dagger)$ 
for any 3 $\times$ 3 matrix $M$.

Applying the factorization of 
$V=v\bar V$ (or $V^\dagger =\bar V^\dagger v^\dagger$) for the gauge function, we find  
\begin{eqnarray}
^tVV&=&\bar V v^\dagger v \bar V=\bar V^2
\end{eqnarray}
and obtain 
\begin{eqnarray}
\Delta R_L &=&{\rm Re Tr}\{{\bar V}^2
\sum_{\mu=1}^4 \sum_{\pm}
 U_{\pm \mu}(s)^tU_{\pm \mu}(s)
\}. 
\end{eqnarray}
This is useful for 
the practical gauge-fixing procedure 
of the maximal SO(3) gauge 
in SU(3) lattice QCD.

\section{SO(3) real algebra method in SU(3) QCD} 

In this section, we present the formalism of the ``SO(3) real algebra method'' for continuum SU(3) QCD at finite baryon-number densities.

Our strategy is to divide 
the SO(3) part, which is essentially real, 
from the SU(3) fermionic determinant 
in the maximal SO(3) gauge.
The SU(3) QCD generating functional 
in the maximal SO(3) gauge 
at finite densities is expressed as 
\begin{widetext}
\begin{eqnarray}
Z^{\rm ~SU(3)}_{\rm SO(3) ~gauge}(\mu)  &=&\int DA e^{-S_G[A]}~\delta^{\rm GF}_{\rm SO(3)}[A]~\Delta_{\rm SO(3)}^{\rm FP}[A]~{\rm Det} (\Slash D+m+\mu\gamma_4) \cr
&=&\int DA e^{-S_G[A]}~\delta^{\rm GF}_{\rm SO(3)}[A]~\Delta_{\rm SO(3)}^{\rm FP}[A]~
{\rm Det} (\Slash D_{\rm SO(3)}+m+\mu\gamma_4) 
\cdot\frac{{\rm Det}(\Slash D+m+\mu\gamma_4)}
{{\rm Det}(\Slash D_{\rm SO(3)}+m+\mu\gamma_4)},
\label{eq:SO(3)Z}
\end{eqnarray}
\end{widetext}
where 
$\delta^{\rm GF}_{\rm SO(3)}[A]$ denotes 
the maximal SO(3) gauge-fixing constraint and $\Delta_{\rm SO(3)}^{\rm FP}[A]$ is the corresponding Faddeev-Popov determinant.
$D_{\rm SO(3)}^\mu$ denotes 
the SO(3) covariant derivative 
defined in Eq.(\ref{eq:SO3CD}), 
$D_{\rm SO(3)}^\mu
\equiv \partial^\mu+igA_{\rm SO(3)}^\mu$.

In Eq.~(\ref{eq:SO(3)Z}), 
the following factor $p[A]$ is generally real, and  
$p[A]$ is non-negative 
for the even-number flavor case ($N_f=2n$) of the same quark mass:
\begin{widetext}
\begin{eqnarray}
p[A] \equiv 
e^{-S_G[A]}~\delta^{\rm GF}_{\rm SO(3)}[A]~\Delta_{\rm SO(3)}^{\rm FP}[A]~
{\rm Det} (\Slash D_{\rm SO(3)}+m+\mu\gamma_4)
&\in& {\bf R} \cr
&\ge& 0 \quad  (N_f=2n).
\end{eqnarray}    
\end{widetext}
Therefore, for the $N_f=2n$ case, this non-negative factor $p[A]$ can be identified as a probability factor in the Monte Carlo method.

Using the relation 
\begin{eqnarray}
D^\mu=D^\mu_{\rm SO(3)}+igA^\mu_{\rm SU(3)/SO(3)}
=D^\mu_{\rm SO(3)}+igA^\mu_{\bar i}T^{\bar i},~~~
\end{eqnarray}
the SU(3) fermionic determinant is expressed as
\begin{eqnarray}
&&{\rm ln~Det}(\Slash D+m+\mu\gamma_4)\cr 
&=&{\rm ln ~Det}(\Slash D_{\rm SO(3)}+m+\mu\gamma_4
+ig\Slash A_{\rm SU(3)/SO(3)})\cr
&=&{\rm Tr~ln} (\Slash D_{\rm SO(3)}+m+\mu\gamma_4
+ig\Slash A_{\rm SU(3)/SO(3)})\cr
&=&{\rm Tr~ln} (\Slash D_{\rm SO(3)}+m+\mu\gamma_4)\cr
&+&{\rm Tr~ln}\left[1+ig\Slash A_{\rm SU(3)/SO(3)}\frac{1}{\Slash D_{\rm SO(3)}+m+\mu\gamma_4}\right].~~~~~
\end{eqnarray}
Then, the fermionic determinant ratio 
in Eq.(\ref{eq:SO(3)Z}) can be written as 
\begin{widetext}
\begin{eqnarray}
&&{\rm ln~}\frac{{\rm Det}(\Slash D+m+\mu\gamma_4)}
{{\rm Det}(\Slash D_{\rm SO(3)}+m+\mu\gamma_4)}
={\rm ln~Det}(\Slash D+m+\mu\gamma_4)
-{\rm ln~Det}(\Slash D_{\rm SO(3)}+m+\mu\gamma_4)\cr 
&=&
{\rm Tr~ln}\left[1+ig\Slash A_{\rm SU(3)/SO(3)}\frac{1}{\Slash D_{\rm SO(3)}+m+\mu\gamma_4}\right]
=-\sum_{n=1}^\infty \frac{(-ig)^{n}}{n}{\rm Tr}\left(
\Slash A_{\rm SU(3)/SO(3)}\frac{1}{\Slash D_{\rm SO(3)}+m+\mu\gamma_4}
\right)^n\cr
&=&\sum_{n=1}^\infty \frac{(-)^{n+1}g^{2n}}{2n}{\rm Tr}\left(
\Slash A_{\rm SU(3)/SO(3)}\frac{1}{\Slash D_{\rm SO(3)}+m+\mu\gamma_4}
\right)^{2n} 
+i\sum_{n=0}^\infty \frac{(-)^ng^{2n+1}}{2n+1}{\rm Tr}\left(
\Slash A_{\rm SU(3)/SO(3)}\frac{1}{\Slash D_{\rm SO(3)}+m+\mu\gamma_4}
\right)^{2n+1}.
\cr
&&
\label{eq:expansion}
\end{eqnarray}
\end{widetext}

This series consists of the SU(3)/SO(3) gluon, $A_{\rm SU(3)/SO(3)}^\mu$, and 
the quark propagator interacting with the SO(3) gluon, 
$A_{\rm SO(3)}^\mu$.
%
Note that the SU(3)/SO(3) gluon $A_{\rm SU(3)/SO(3)}^\mu$ 
is real, 
\begin{eqnarray}
A_{\rm SU(3)/SO(3)}^\mu(x)^*=A_{\rm SU(3)/SO(3)}^\mu(x),
\end{eqnarray}
and the SO(3) covariant derivative 
$D_{\rm SO(3)}^\mu$ is also real,
\begin{eqnarray}
{D_{\rm SO(3)}^\mu}^*=(\partial^\mu+igA_{\rm SO(3)}^\mu)^*=D_{\rm SO(3)}^\mu,
\end{eqnarray}
because of
${A_{\rm SO(3)}^\mu}^*=-A_{\rm SO(3)}^\mu$.
Then, we find for real $\mu$   
\begin{eqnarray}
   && \left[{\rm Tr}\left(
\Slash A_{\rm SU(3)/SO(3)}\frac{1}{\Slash D_{\rm SO(3)}+m+\mu\gamma_4}
\right)^n\right]^* \cr
&=&
{\rm Tr}\left(C
\Slash A_{\rm SU(3)/SO(3)}\frac{1}{\Slash D_{\rm SO(3)}+m+\mu\gamma_4}
C^{-1} \right)^n \cr
&=&
{\rm Tr}\left(
\Slash A_{\rm SU(3)/SO(3)}\frac{1}{\Slash D_{\rm SO(3)}+m+\mu\gamma_4}
\right)^n \in {\bf R},
\label{eq:expansion_real}
\end{eqnarray}
where $C$ is the complex  conjugation matrix satisfying 
$\gamma_\mu^*=C \gamma_\mu C^{-1}$,
i.e., Eq.~(\ref{eq:CC}).
Therefore, the fermionic determinant ratio is expressed as
\begin{eqnarray}
{\rm ln~}\frac{{\rm Det}(\Slash D+m+\mu\gamma_4)}
{{\rm Det}(\Slash D_{\rm SO(3)}+m+\mu\gamma_4)}&=&p(\mu)+iq(\mu), \cr
{\rm i.e.,} \quad \frac{{\rm Det}(\Slash D+m+\mu\gamma_4)}
{{\rm Det}(\Slash D_{\rm SO(3)}+m+\mu\gamma_4)}&=&
e^{p(\mu)+iq(\mu)}, 
\end{eqnarray}
where $p(\mu)$ and 
$q(\mu)$ are the real numbers 
defined by 
\begin{widetext}
\begin{eqnarray}
&&p(\mu)\equiv \sum_{n=1}^\infty \frac{(-)^{n+1}g^{2n}}{2n}{\rm Tr}\left(
\Slash A_{\rm SU(3)/SO(3)}\frac{1}{\Slash D_{\rm SO(3)}+m+\mu\gamma_4}
\right)^{2n} \in {\bf R}, \cr
&&q(\mu)\equiv \sum_{n=0}^\infty \frac{(-)^ng^{2n+1}}{2n+1}{\rm Tr}\left(
\Slash A_{\rm SU(3)/SO(3)}\frac{1}{\Slash D_{\rm SO(3)}+m+\mu\gamma_4}
\right)^{2n+1} \in {\bf R}.
\end{eqnarray}
\end{widetext}
For the fermionic determinant ratio, 
the real number $p(\mu)$ yields a harmless real factor of $e^{p(\mu)}$, and 
$q(\mu)$ gives the phase factor $e^{iq(\mu)}$,  
which has the potential 
to cause the sign problem in the Monte Carlo method.

Using the real nature of Eq.~(\ref{eq:expansion_real}) 
and the anti-hermitian nature of 
$\Slash D_{\rm SO(3)}$, 
we find
\begin{eqnarray}
   && {\rm Tr}\left(
\Slash A_{\rm SU(3)/SO(3)}\frac{1}{\Slash D_{\rm SO(3)}+m-\mu\gamma_4}
\right)^n \cr
&=&
{\rm Tr}\left[\left(
\Slash A_{\rm SU(3)/SO(3)}\frac{1}{\Slash D_{\rm SO(3)}+m-\mu\gamma_4}
\right)^n\right]^\dagger \cr
&=&
{\rm Tr}\left(
\Slash A_{\rm SU(3)/SO(3)}\frac{1}{-\Slash D_{\rm SO(3)}+m-\mu\gamma_4}
\right)^n \cr
&=&
{\rm Tr}\left[\gamma_5\left(
\Slash A_{\rm SU(3)/SO(3)}\frac{1}{-\Slash D_{\rm SO(3)}+m-\mu\gamma_4}
\right)^n\gamma_5\right] \cr
&=& (-)^n~{\rm Tr}\left(
\Slash A_{\rm SU(3)/SO(3)}\frac{1}{\Slash D_{\rm SO(3)}+m+\mu\gamma_4}
\right)^n.
\label{eq:sym}
\end{eqnarray}
Then, $p(\mu)$ and $q(\mu)$ 
are even and odd functions of $\mu$, 
respectively, 
\begin{eqnarray}
p(-\mu)=p(\mu), \quad 
q(-\mu)=-q(\mu).
\label{eq:even-odd}
\end{eqnarray}
This leads to 
\begin{eqnarray}
q(\mu=0)=0,   
\end{eqnarray}
meaning that 
the fermionic determinant ratio 
is real 
at zero density, $\mu=0$.
Using the relation,
\begin{eqnarray}
&&\frac{1}{\Slash D_{\rm SO(3)}+m+\mu\gamma_4}\cr
&=&
\frac{1}{\Slash D_{\rm SO(3)}+m}
+\frac{1}{\Slash D_{\rm SO(3)}+m}
(-\mu\gamma_4)
\frac{1}{\Slash D_{\rm SO(3)}+m+\mu\gamma_4}\cr
&=&\frac{1}{\Slash D_{\rm SO(3)}+m}
\sum_{k=0}^\infty\left(-\mu\gamma_4
\frac{1}{\Slash D_{\rm SO(3)}+m}\right)^k,
\end{eqnarray}
further expansion of the series is possible for both $p(\mu)$ and $q(\mu)$, 
\begin{widetext}
\begin{eqnarray}
&&p(\mu)\equiv \sum_{n=1}^\infty \frac{(-)^{n+1}g^{2n}}{2n}{\rm Tr}\left[
\Slash A_{\rm SU(3)/SO(3)}\frac{1}{\Slash D_{\rm SO(3)}+m}\sum_{k=0}^\infty\left(-\mu\gamma_4
\frac{1}{\Slash D_{\rm SO(3)}+m}\right)^k
\right]^{2n} \in {\bf R}, \cr
&&q(\mu)\equiv \sum_{n=0}^\infty \frac{(-)^{n+1}g^{2n+1}}{2n+1}{\rm Tr}\left[
\Slash A_{\rm SU(3)/SO(3)}\frac{1}{\Slash D_{\rm SO(3)}+m}\sum_{k=0}^\infty\left(\mu\gamma_4
\frac{1}{\Slash D_{\rm SO(3)}+m}\right)^{2k+1}
\right]^{2n+1} \in {\bf R}.
\end{eqnarray}
\end{widetext}
The series for even $k$ vanishes in the second equation due to a symmetric relation similar to Eq.~(\ref{eq:sym}).
The second equation again shows 
that $q(\mu)$ is an odd function of $\mu$.
This expansion is useful when $\mu$ is small, yet preserves the effect of $\mu$ in the SO(3) fermion determinant.

If the norm of $A^\mu_{\rm SU(3)/SO(3)}$ is well suppressed in the maximally SO(3) gauge, 
$q(\mu)$ may not fluctuate significantly over the sampled gauge configurations, 
and also the infinite series in Eq.~(\ref{eq:expansion}) could be truncated to a finite one.
Even after the truncation, 
the effect of the chemical potential $\mu$ 
is firmly included in the SO(3) fermionic determinant. 

\section{SO(3) real algebra method \\ in SU(3) lattice QCD}

Now, we formulate the SO(3) real algebra method in SU(3) lattice QCD 
at a finite density with 
a quark-number chemical potential $\mu$.
Using the SU(3) lattice fermion kernel $K[U;\mu]$,
the SU(3) fermionic determinant is expressed as 
${\rm Det}~K[U;\mu]$.
The SU(3) lattice QCD generating functional 
in the maximal SO(3) gauge at finite densities 
is expressed as 
\begin{widetext}
\begin{eqnarray}
Z^{\rm ~SU(3)}_{\rm SO(3)~gauge}(\mu) &=&\int DU e^{-S_G[U]}~\delta^{\rm GF}_{\rm SO(3)}[U]~\Delta_{\rm SO(3)}^{\rm FP}[U]~{\rm Det}~K[U;\mu] \cr
&=&\int DU e^{-S_G[U]}~\delta^{\rm GF}_{\rm SO(3)}[U]~\Delta_{\rm SO(3)}^{\rm FP}[U]~
{\rm Det}~K[u;\mu]
\cdot\frac{{\rm Det}~K[U;\mu]}
{{\rm Det}~K[u;\mu]},
\label{eq:LSO(3)Z}
\end{eqnarray}
\end{widetext}
where 
$\delta^{\rm GF}_{\rm SO(3)}[U]$ denotes 
the maximal SO(3) gauge-fixing constraint and $\Delta_{\rm SO(3)}^{\rm FP}[U]$ is the corresponding Faddeev-Popov determinant 
in lattice QCD.

In Eq.~(\ref{eq:LSO(3)Z}), 
the following factor $p[U]$ is generally real, and  
$p[U]$ is non-negative 
for the even-number flavor case ($N_f=2n$) of the same quark mass:
\begin{widetext}
\begin{eqnarray}
p[U] \equiv 
e^{-S_G[U]}~\delta^{\rm GF}_{\rm SO(3)}[U]~\Delta_{\rm SO(3)}^{\rm FP}[U]
~{\rm Det}~K[u;\mu]
&\in& {\bf R} \cr
&\ge& 0 \quad  (N_f=2n).
\end{eqnarray}    
\end{widetext}
Therefore, for the $N_f=2n$ case, 
the non-negative factor $p[U]$ can be identified as a probability factor.

The SO(3) real algebra method employs the non-negative factor 
$p[U]$ as a probability factor of Monte Carlo calculations for the $N_f=2n$ case with the same quark mass.
In this method,  
SU(3) gauge configurations 
$\{U_k\}_{k=1,2,...,N}$ can be generated 
by the most importance sampling 
with the weight of $p[U]$. 
This procedure is achieved through the Monte Carlo update 
for the SU(3) action factor $e^{-S_G[U]}$ and 
the SO(3) fermionic determinant ${\rm Det}~K[u;\mu]$,   
while applying the maximal SO(3) gauge fixing. 
In fact, the SO(3) real algebra method alternates between Monte Carlo updates and the maximal SO(3) gauge fixing.

For an arbitrary operator  $O[U]$, 
its expectation value is 
evaluated as 
\begin{eqnarray}
\langle O[U]\rangle_{\rm SU(3)}
&=&\left\langle O[U]\cdot 
\frac{{\rm Det}~K[U;\mu]}
{{\rm Det}~K[u;\mu]}
\right\rangle_{\rm SO(3)} \cr
&=&\sum_{k=1}^N O[U_k]\cdot 
\frac{{\rm Det}~K[U_k;\mu]}
{{\rm Det}~K[u_k;\mu]},
\end{eqnarray}
where $U_k=M_ku_k$ denotes the $k$~th gauge configuration.
Here, $\langle...\rangle_{\rm SO(3)}$ means the 
expectation value with 
the weight $p[U]$, 
which is real and non-negative  
for the case of $N_f=2n$ 
with the same quark mass.
In fact, the SO(3) real algebra method is a 
type of reweighting method \cite{FS88,FS89,SOK24}.
Note that this method only uses gauge fixing during the most importance  sampling, and the final result is completely gauge invariant.

The SO(3) real algebra method can be 
easily extended to finite temperature QCD 
by imposing temporal (anti)periodicity 
\cite{R12}. 
In fact, this method can be used for  SU(3) lattice QCD at finite densities and temperatures.

For arbitrary hermite operators, 
the expectation value 
$\langle O[U] \rangle_{\rm SU(3)}$
itself takes a real physical value.
The problem is that there could be 
large fluctuations in the phase 
factor of the determinant ratio, 
which would cause numerical 
Monte Carlo simulations to fail.

However, if one can perform good sampling and suppress the fluctuation of the phase factor  in each sampled configuration, 
the Monte Carlo calculation will be possible.
Although the SU(3) fermionic determinant 
${\rm Det}~K[U_k;\mu]$ is gauge invariant, its numerical evaluation 
depends on the sampled 
gauge configurations, 
and there are possibilities for improvement in the sampling process.

In fact, if the phase factor of 
the determinant ratio  
${\rm Det}~K[U_k;\mu]/
{\rm Det}~K[u_k;\mu]$
is not significantly fluctuating 
in the Monte Carlo sampled gauge configurations, then lattice QCD calculations at finite densities would be feasible for the $N_f=2n$ case.

On this point, in the maximal SO(3) gauge, 
the SU(3) gauge configurations 
are expected to be maximally close to 
the SO(3) ones, and hence the phase factor of 
the determinant ratio  
${\rm Det}~K[U_k;\mu]/
{\rm Det}~K[u_k;\mu]$
in the sampled gauge configurations 
may not fluctuate significantly 
during the most importance sampling
in the Monte Carlo method.

The next important step is 
to check the phase factor of 
the determinant ratio  
${\rm Det}~K[U_k;\mu]/
{\rm Det}~K[u_k;\mu]$
in the maximal SO(3) gauge 
in an actual lattice QCD Monte Carlo simulation.
The degree of fluctuation would depend on 
the lattice volume, the chemical potential $\mu$, 
the fermion mass $m$ and so on.
If there is a window that is close to the physical situation, this approach would be 
realistic for exploring finite density SU(3) QCD.

\section{Summary and conclusion}

In this paper, 
we have proposed the SO(3) real algebra method
for SU(3) lattice QCD calculations at finite baryon-number densities.
In this method, we have divided the SU(3) gauge variable into the SO(3) and SU(3)/SO(3) parts. 
Then, we have introduced the maximal SO(3) gauge by minimizing the SU(3)/SO(3) part of the SU(3) gauge variable. 
We have considered 
the SO(3) fermionic determinant, 
that is, the fermionic determinant 
of the SO(3) part of the SU(3) gauge variable.

We have noted that the SO(3) fermionic determinant is real, and it is non-negative for the even-number flavor case ($N_f=2n$) of the same quark mass, 
e.g., $m_u=m_d$.
In the Monte Carlo calculation, 
as well as the positive 
SU(3) gauge action factor $e^{-S_G}$, 
the SO(3) real algebra method has used  the SO(3) fermionic determinant in the maximal SO(3) gauge for the $N_f = 2n$ case. 

The SO(3) real algebra method alternates between 
the maximal SO(3) gauge fixing and 
Monte Carlo updates on the SO(3) determinant and $e^{-S_G}$.
%
%
After the most importance sampling, we have treated 
the remaining ratio of the SU(3) and SO(3) fermionic determinants as a weight factor in the reweighting method. 
In fact, we have only used gauge fixing for the most importance sampling, 
and the final result is fully gauge invariant.
If the phase factor of the ratio does 
not fluctuate significantly in the sampled gauge configurations, then 
SU(3) lattice QCD calculations at finite densities would be feasible.
This method can be extended
to finite temperature QCD by imposing temporal
(anti)periodicity. In fact, this method can be used to
study QCD at finite densities and temperatures.

We summarize the lattice QCD procedure of the SO(3) real algebra method for finite density QCD:
\begin{enumerate}
\item 
Perform maximal SO(3) gauge fixing 
to maximize $\sum_{\mu=1}^4{\rm Re Tr}\{U_\mu(s)^t U_\mu(s)\}$.
\item
Perform a Monte Carlo calculation 
for the SU(3) action factor $e^{-S_G[U]}$ and 
the SO(3) fermionic determinant ${\rm Det}~K[u;\mu]$.
\item
Repeat steps 1 and 2 alternately until thermalization is achieved. Then, sample the SU(3) gauge configurations 
$\{U_k\}_{k=1,2,...,N}$.
\item
Use the sampled SU(3) gauge configurations in the above and 
include the fermionic determinant ratio 
${\rm Det}~K[U;\mu]/{\rm Det}~K[u;\mu]$
as a weight factor 
when calculating the expectation value.
\end{enumerate}

%

The remaining serious problem is the possibility of  significant fluctuations in the phase factor of the fermionic determinant ratio, 
although the SU(3) gauge variable is expected to 
closely resemble its SO(3) part 
in the maximal SO(3) gauge.
As a next important work, 
we will investigate the 
the phase factor distribution 
of the fermionic determinant ratio 
in the Monte Carlo sampled configurations  
using the SO(3) real algebra method. 
The degree of fluctuation would 
depend on the lattice volume, 
the chemical potential $\mu$, the temperature $T$,  
the fermion mass $m$ and so on. 

If there is a window that is close to the physical situation, this approach would be realistic for  
lattice calculations of 
SU(3) QCD at finite densities and temperatures, and the QCD phase diagram would be obtained.

The series in Eq.~(\ref{eq:expansion}) may be useful.
Note again that this series consists of the SU(3)/SO(3) gluon field and the quark propagator interacting 
with the SO(3) gluon, which are both essentially real in the trace.

Compared to other methods, 
the SO(3) real algebra method is a type of reweighting method \cite{FS88,FS89,SOK24}.
This method is philosophically similar to the
Lefschetz thimble method \cite{W11,ABB16,MKO17,MKO18,FU17,FM21}, which reduces phase fluctuations by taking a suitable complex path, while this method reduces them by taking a suitable gauge.
Unlike the Taylor expansion method \cite{FP02,AEHKK02}, 
this method does not use an expansion by $\mu$, 
so its applicable region is not limited to small $\mu$.
Even if this method alone may not sufficiently suppress phase fluctuations, combining it with the other methods mentioned in Sec.~I may enable the numerical simulation of finite density SU(3) QCD feasible.

Finally, we briefly provide some related perspectives on the maximal SO(3) real algebra method and the maximal SO(3) gauge.
First, the generalization of this method to SU($N$) gauge theories is straightforward by performing a similar procedure that divides the SU($N$) gauge variable into SO($N$) and SU($N$)/SO($N$) parts.
Second, it is interesting to investigate
``SO(3) dominance'' 
in the maximal SO(3) gauge,
motivated by Abelian dominance \cite{IS00,SY90,IS99,SS14,SS15,t81,EI82,AS99} 
in the maximally Abelian gauge.

\section*{Acknowledgements}
We thank Dr. T. M. Doi for his useful advice on the direct calculation of the fermionic determinant. 
K.T. is supported by JSPS Research Fellowship for Young Scientists.


\begin{thebibliography}{99} 

\bibitem{GW73}
D. J. Gross and F. Wilczek, 
{\it Ultraviolet Behavior of Non-Abelian Gauge Theories}, 
Phys. Rev. Lett. {\bf 30}, 1343-1346 (1973);
{\it Asymptotically Free Gauge Theories I}, 
Phys. Rev. D{\bf 8}, 3633-3652 (1973);
{\it Asymptotically Free Gauge Theories II}, 
Phys. Rev. D{\bf 9}, 980-993, (1974).

\bibitem{P73}
H. D. Politzer, 
{\it Reliable Perturbative Results for Strong Interactions?},
Phys. Rev. Lett. {\bf 30}, 1346-1349 (1973).

\bibitem{W24}
F. Wilczek, 
{\it QCD at 50: Golden Anniversary, Golden Insights, Golden Opportunities}, 
in ``2023 Majorana Summer School'', Erice, 2023,
arXiv:2403.06038 [hep-ph] (2024) and its references.

\bibitem{M09}
T. Muta, {\it Foundations Of Quantum Chromodynamics: An Introduction to Perturbative Methods in Gauge Theories}, (World Scientific, 2019) and its references.

\bibitem{BP69}
J.D. Bjorken and E.A. Paschos, 
{\it Inelastic Electron-Proton and $\gamma$-Proton Scattering and the Structure of the Nucleon}, 
Phys. Rev. {\bf 185}, 1975-1982 (1969).

\bibitem{S23}
For a recent review article, 
H. Suganuma,
{\it Quantum Chromodynamics, Quark Confinement and Chiral Symmetry Breaking: a Bridge Between Elementary Particle Physics and Nuclear Physics},
``Handbook of Nuclear Physics", (Springer, 2023) 22-1. 

\bibitem{NJL61}
Y. Nambu and G. Jona-Lasinio, 
{\it Dynamical Model of Elementary Particles Based on an Analogy with Superconductivity. I}, 
Phys. Rev. {\bf 122}, 345-358 (1961);
{\it Dynamical Model of Elementary Particles Based on an Analogy with Superconductivity. II},
Phys. Rev. {\bf 124}, 246-254 (1961).

\bibitem{W74} 
K.G. Wilson, 
{\it Confinement of quarks}, 
Phys. Rev. D{\bf 10}, 2445-2459 (1974).

\bibitem{KS75}
J. B. Kogut and L. Susskind, 
{\it Hamiltonian formulation of Wilson's lattice gauge theories}, 
Phys. Rev. D{\bf 11}, 395-408 (1975).

\bibitem{C79} 
M. Creutz, 
{\it Confinement and the Critical Dimensionality of Space-Time},
Phys. Rev. Lett. {\bf 43}, 553-556 (1979), Erratum {\it ibid.} {\bf 43}, 890 (1979).

\bibitem{C80} 
M. Creutz, 
{\it Monte Carlo study of quantized SU(2) gauge theory},
Phys. Rev. D{\bf 21}, 2308-2315 (1980).

\bibitem{DKPR87}
S. Duane, A.D. Kennedy, B.J. Pendleton, and 
D. Roweth,
{\it Hybrid Monte Carlo},
Phys. Lett. {\bf B195}, 216-222 (1987).

\bibitem{R12}
H. J. Rothe,
{\it Lattice Gauge Theories, 4th Edition}, 
(World Scientific, 2012), 
and its references.

\bibitem{HOTQCD14}
HotQCD Collaboration (A. Bazavov et al.), 
{\it Equation of state in (2+1)-flavor QCD}
Phys. Rev. D{\it 90}, 094503 (2014) 
and its references.

\bibitem{N22}
For a recent review on the sign problem, 
K. Nagata, 
{\it Finite-density lattice QCD and sign problem: Current status and open problems},
Prog. Part. Nucl. Phys. {\bf 127}, 
103991 (2022).

\bibitem{YHM05}
K. Yagi, T. Hatsuda and Y. Miake,
{\it Quark-Gluon Plasma: From Big Bang to Little Bang}, 
(Cambridge University Press, 2005) and its references.

\bibitem{M73}
A.B. Migdal, {\it $\pi$ condensation in nuclear matter},
Phys. Rev. Lett. {\bf 31}, 257-260 (1973).

\bibitem{TTTT78}
T. Takatsuka, K. Tamiya, T. Tatsumi, and 
R. Tamagaki, 
{\it Solidification and Pion Condensation in Nuclear Medium. Alternating Layer Spin Structure with One-Dimensional Localization Accompanying $\pi^0$ Condensate}, 
Prog. Theor. Phys. {\bf 59}, 1933-1955 (1978).

\bibitem{ASRS08}
M.G. Alford, A. Schmitt,
K. Rajagopal, and T. Sch\"afer, 
{\it Color superconductivity in dense quark matter},
Rev. Mod. Phys. {\bf 80}, 1455-1515 (2008) and its references.

\bibitem{MP07}
L. McLerran and R.D. Pisarski,
{\it Phases of dense quarks at large $N_c$},
Nucl. Phys. {\bf A796}, 83-100 (2007).

\bibitem{MRS09}
L. McLerran, K. Redlich, and C. Sasaki, 
{\it Quarkyonic Matter and Chiral Symmetry Breaking},
Nucl. Phys. {\bf A824}, 86-100 (2009).


\bibitem{FP02}
P. de Forcrand and O. Philipsen, 
{\it The QCD phase diagram for small densities from imaginary chemical potential}
Nucl. Phys. B{\bf 642}, 290-306 (2002).

\bibitem{AEHKK02}
C.R. Allton, S. Ejiri, S.J. Hands, O.~Kaczmarek, 
F. Karsch et al., 
{\it QCD thermal phase transition in the presence of a small chemical potential}, 
Phys. Rev. D{\bf 66}, 074507 (2002).

\bibitem{PW81}
G. Parisi and Y. Wu, 
{\it Perturbation Theory Without Gauge Fixing}, 
Sci. Sin. {\bf 24}, 483 (1981).

\bibitem{P83}
G. Parisi, 
{\it On Complex Probabilities}, 
Phys. Lett. {\bf 131B}, 393-395 (1983).

\bibitem{ASS10}
G. Aarts, E. Seiler, and I.-O. Stamatescu, 
{\it The Complex Langevin method: When can it be trusted?}, 
Phys. Rev. D{\bf 81}, 054508 (2010).

\bibitem{S14}
D. Sexty, 
{\it Simulating full QCD at nonzero density using the complex Langevin equation},
Phys. Lett. {\bf B729}, 108-111 (2014).

\bibitem{NNS16}
K. Nagata, J. Nishimura, and S. Shimasaki,
{\it Argument for justification of the complex Langevin method and the condition for correct convergence}, 
Phys. Rev. D{\bf 94}, 114515 (2016).

\bibitem{TD16}
S. Tsutsui and T.M. Doi,
{\it Improvement in complex Langevin dynamics from a view point of Lefschetz thimbles}, 
Phys. Rev. D{\bf 94}, 074009 (2016).

\bibitem{FS88}
A.M. Ferrenberg and R.H. Swendsen, 
{\it New Monte Carlo technique for studying phase transitions},
Phys. Rev. Lett. {\bf 61}, 2635 (1988), 
Erratum {\it ibid.} {\bf 63}, 1658 (1989).

\bibitem{FS89}
A.M. Ferrenberg and R.H. Swendsen, 
{\it Optimized Monte Carlo data analysis},
Phys. Rev. Lett. {\bf 63}, 1195 (1989).

\bibitem{SOK24}
H. Suganuma, H. Ohata and M. Kitazawa,
{\it Thermodynamic Potential of the Polyakov Loop in SU(3) Quenched Lattice QCD}, 
Int. J. Mod. Phys. {\bf A39}, 2443012 (2024).

\bibitem{FK02}
Z. Fodor and S.D. Katz, 
{\it A new method to study lattice QCD at finite temperature and chemical potential}, 
Phys. Lett. {\bf B534}, 87-92 (2002).

\bibitem{AKW99}
M. Alford, A. Kapustin, and F. Wilczek, 
{\it Imaginary chemical potential and finite fermion density on the lattice}, 
Phys. Rev. D{\bf 59}, 054502 (1999).

\bibitem{DL03}
M. D'Elia and M.-P. Lombardo, 
{\it Finite density QCD via imaginary chemical potential}, 
Phys. Rev. D{\bf 67}, 014505 (2003). 

\bibitem{NN11}
K. Nagata and A. Nakamura, 
{\it Imaginary Chemical Potential Approach for the Pseudo-Critical Line in the QCD Phase Diagram with Clover-Improved Wilson Fermions}
Phys. Rev. D{\bf 83}, 114507 (2011).

\bibitem{BBFGKRS15}
R. Bellwied, S. Borsanyi, Z. Fodor, J. Guenther, S.D. Katz, C. Ratti, and K.K. Szabo, 
{\it The QCD phase diagram from analytic continuation},
Phys. Lett. B{\bf 751}, 559-564 (2015).

\bibitem{W11}
E. Witten, 
{\it Analytic Continuation Of Chern-Simons Theory}, 
Adv. Math. {\bf 50}, 347 (2011).

\bibitem{ABB16}
A. Alexandru, G. Basar, and P. Bedaque, 
{\it Monte Carlo algorithm for simulating fermions on Lefschetz thimbles}, 
Phys. Rev. D{\bf 93}, 014504 (2016).

\bibitem{MKO17}
Y. Mori, K. Kashiwa, and A. Ohnishi, 
{\it Toward solving the sign problem with path optimization method}
Phys. Rev. D{\bf 96} 111501 (2017).

\bibitem{MKO18}
Y. Mori, K. Kashiwa, and A. Ohnishi,
{\it Application of a neural network to the sign problem via the path optimization method},
PTEP {\bf 2018}, 023B04 (2018).

\bibitem{FU17}
M. Fukuma and N. Umeda,
{\it Parallel tempering algorithm for integration over Lefschetz thimbles},
PTEP {\bf 2017}, 073B01 (2017).

\bibitem{FM21}
M. Fukuma and N. Matsumoto,
{\it Worldvolume approach to the tempered Lefschetz thimble method},
PTEP {\bf 2021}, 023B08 (2021).

\bibitem{FLLP12}
M. Fromm, J. Langelage, S. Lottini, and O. Philipsen, 
{\it The QCD deconfinement transition for heavy quarks and all baryon chemical potentials},
JHEP {\bf 01}, 042 (2012).

\bibitem{HMY12}
M. Hanada, Y. Matsuo, and N. Yamamoto,
{\it Sign problem and phase quenching in finite-density QCD: Models, holography, and lattice},
Phys. Rev. D{\bf 86}, 074510 (2012).

\bibitem{BGIKM15}
V.V. Braguta, V.A. Goy, E.-M. Ilgenfritz,
A.Yu. Kotov, A.V. Molochkov et al.,
{\it Two-Color QCD with Non-zero Chiral Chemical Potential}, 
JHEP {\bf 06} 094 (2015) 094.

\bibitem{HKLM99}
S. Hands, J.B. Kogut, M.-P. Lombardo,
and S.E. Morrison, 
{\it Symmetries and spectrum of SU(2) lattice gauge theory at finite chemical potential},
Nucl. Phys. {\bf B558}, 327-346 (1999).

\bibitem{BBIKM18}
V.G. Bornyakov, V.V. Braguta, E. -M. Ilgenfritz,
A. Yu. Kotov, A.V. Molochkov et al., 
{\it Observation of deconfinement in a cold dense quark medium},
JHEP {\bf 03}, 161 (2018).

\bibitem{IIL20}
K. Iida, E. Itou, and T.-G. Lee, 
{\it Two-colour QCD phases and the topology at low temperature and high density}
JHEP{\bf 01}, 181 (2020).

\bibitem{I25}
For a recent review on dense two-color QCD, 
E. Itou, 
{\it Lattice results for the equation of state in dense
QCD-like theories}, arXiv:2508.03090 [hep-lat] (2025).

\bibitem{SS01}
D.T. Son and M.A. Stephanov, 
{\it QCD at finite isospin density}
Phys. Rev. Lett. {\bf 86} (2001) 592-595 (2001).

\bibitem{KS02}
J.B. Kogut and D.K. Sinclair, 
{\it Quenched lattice QCD at finite isospin density and related theories},
Phys. Rev. D{\bf 66}, 014508 (2002).

\bibitem{HOTQCD19}
Hot QCD Collaboration (A. Bazavov et al.),
{\it Chiral crossover in QCD at zero and non-zero chemical potentials},
Phys. Lett. B{\bf 795}, 15-21 (2019).

\bibitem{Y11}
A. Yamamoto, 
{\it Chiral magnetic effect in lattice QCD with a chiral chemical potential}, 
Phys. Rev. Lett. {\bf 107}, 031601 (2011).

\bibitem{CME08}
K. Fukushima, D.E. Kharzeev, and H.J. Warringa,
{\it The Chiral Magnetic Effect},
Phys. Rev. D{\bf 78}, 074033 (2008).

\bibitem{ST91}
H. Suganuma and T. Tatsumi,
{\it On the Behavior of Symmetry and Phase Transitions in a Strong Electromagnetic Field}, 
Ann. Phys. {\bf 208}, 470-508 (1991).

\bibitem{MS15}
V.A. Miransky and I.A. Shovkovy, 
{\it Quantum field theory in a magnetic field: From quantum chromodynamics to graphene and Dirac semimetals}
Phys. Rept. {\bf 576}, 1-209 (2015) 1-209.

\bibitem{DMS10}
M. D'Elia, S. Mukherjee, and F. Sanfilipp, 
{\it QCD Phase Transition in a Strong Magnetic Background}, Phys. Rev. {\bf D82}, 051501 (2010).

\bibitem{BBLRS84}
A.P. Balachandran, A. Barducci, F. Lizzi, V.G.J. Rodgers, and A. Stern, 
{\it A Doubly Strange Dibaryon in the Chiral Model}
Phys. Rev. Lett. {\bf 52}, 887-890 (1984).

\bibitem{BLRS85}
A.P. Balachandran, F. Lizzi, V.G.J. Rodgers, and A. Stern, 
{\it Dibaryons as Chiral Solitons}, 
Nucl. Phys. {\bf B256}, 525-556 (1985).

\bibitem{JK85}
R.L. Jaffe and C.L. Korpa, 
{\it Semiclassical quantization of the dibaryon skyrmion}, 
Nucl. Phys. {\bf B258}, 468-482 (1985).

\bibitem{SS98}
T. Sakai and H. Suganuma, 
{\it H dibaryon matter in the Skyrme model on a hypersphere}, 
Phys. Lett. {\bf B430}, 168-173 (1998).

\bibitem{MNS17}
K. Matsumoto, Y. Nakagawa, and H. Suganuma, 
{\it A Study of the H-dibaryon in Holographic QCD},
JPS Conf. Proc. {\bf 13}, 020014 (2017).

\bibitem{SM17}
H. Suganuma and K. Matsumoto, 
{\it Holographic QCD for H-dibaryon (uuddss)},
EPJ Web Conf. {\bf 137}, 13018 (2017).

\bibitem{IS00}
H. Ichie and H. Suganuma, 
{\it Monopoles and gluon fields in QCD in the maximally Abelian gauge},
Nucl. Phys. {\bf B574}, 70-106 (2000).

\bibitem{JJP85}
A. Jackson, A.D. Jackson, and V. Pasquier,
{\it The Skyrmion-Skyrmion Interaction},
Nucl. Phys. {\bf A432}, 567-609 (1985).

\bibitem{YA85}
H. Yabu and  K. Ando, 
{\it Static N-N and N-N Interactions in the Skyrme Model}, 
Prog. Theor. Phys. {\bf 74}, 750–764 (1985).

\bibitem{KLSW87}
A.S. Kronfeld, M.L. Laursen, G. Schierholz, and U.J. Wiese, 
{\it Monopole Condensation and Color Confinement},
Phys. Lett. {\bf B198}, 516-520 (1987).

\bibitem{KSW87}
A.S. Kronfeld, G. Schierholz, and 
U.J. Wiese,
{\it Topology and Dynamics of the Confinement Mechanism}, 
Nucl. Phys. {\bf B293}, 461-478
(1987).

\bibitem{SY90}
T. Suzuki and I. Yotsuyanagi,
{\it A possible evidence for Abelian dominance in quark confinement},
Phys. Rev. D{\bf 42}, 4257-4260 (1990).

\bibitem{IS99}
H. Ichie and H. Suganuma, 
{\it Abelian dominance for confinement and random phase property of off diagonal gluons in the maximally Abelian gauge},
Nucl. Phys. {\bf B548}, 365-382 (1999).



\bibitem{SS14}
N. Sakumichi and H. Suganuma, 
{\it Perfect Abelian Dominance of Quark Confinement in SU(3) QCD}, 
Phys. Rev. D{\bf 90}, 111501(R) (2014).

\bibitem{SS15}
N. Sakumichi and H. Suganuma, 
{\it Three-Quark Potential and Abelian Dominance of Confinement in SU(3) QCD}, 
Phys. Rev. D{\bf 92}, 034511 (2015).


\bibitem{t81}
G.~'t~Hooft, 
{\it Topology of the Gauge Condition and New Confinement Phases in Nonabelian Gauge Theories},
Nucl. Phys. {\bf B190}, 455-478 (1981).

\bibitem{EI82}
Z.F. Ezawa and A. Iwazaki, 
{\it Abelian Dominance and Quark Confinement in Yang-Mills Theories},
Phys. Rev. D{\bf 25}, 2681-2689 (1982).

\bibitem{AS99}
K. Amemiya and H. Suganuma, 
{\it Off-diagonal Gluon Mass Generation and Infrared Abelian Dominance in the Maximally Abelian Gauge in Lattice QCD}, 
Phys. Rev. D{\bf 60}, 114509 (1999).


\end{thebibliography}
\end{document}